\documentclass{article}
\usepackage{graphicx} 
\usepackage{booktabs} 
\usepackage{array}    
\usepackage{ragged2e} 
\usepackage{amsmath}   
\usepackage{amssymb}   
\usepackage{natbib} 
\usepackage[utf8]{inputenc}
\usepackage[a4paper,%
            left=2.5cm,    
            right=2.5cm,   
            top=2.5cm,     
            bottom=2.5cm   
           ]{geometry}
\usepackage{float} 
\usepackage[hidelinks]{hyperref}
\usepackage{enumitem}
\setlist[enumerate]{leftmargin=1.2em}

\usepackage{indentfirst}
\usepackage{xcolor}

\newcolumntype{L}[1]{>{\raggedright\arraybackslash}p{#1}}
\usepackage{booktabs}
\usepackage{amsmath}
\usepackage{wrapfig}
\usepackage{url}
\title{\textbf{An Efficient Calibration Framework for Volatility Derivatives under Rough Volatility with Jumps}}

\author{%
  Keyuan Wu \footnote{Department of Mathematics, University of Southern California. Email: \texttt{<keyuanwu@usc.edu>}.} \and
  Tenghan Zhong \footnote{Department of Mathematics, University of Southern California. Email: \texttt{<tenghanz@usc.edu>}.} \and
  Yuxuan Ouyang \footnote{Department of Mathematics, University of Southern California. Email: \texttt{<yuxuanou@usc.edu>}.}
}
\date{2025} 

\begin{document}

\maketitle
\setcounter{footnote}{0}\renewcommand{\thefootnote}{\arabic{footnote}}
\begin{abstract} 
We present a fast and robust calibration method for stochastic volatility models that admit Fourier-analytic transform-based pricing via characteristic functions. The design is structure-preserving: we keep the original pricing transform and (i) split the pricing formula into data-independent integrals and a market-dependent remainder; (ii) precompute those data-independent integrals with GPU acceleration; and (iii) approximate only the remaining, market-dependent pricing map with a small neural network. We instantiate the workflow on a rough volatility model with tempered-stable jumps tailored to power-type volatility derivatives and calibrate it to VIX options with a global-to-local search. We verify that a pure-jump rough volatility model adequately captures the VIX dynamics, consistent with prior empirical findings, and demonstrate that our calibration method achieves high accuracy and speed.
\end{abstract}


 \newpage

\tableofcontents
\newpage
\section{Introduction}
\subsection{Literature Review}
\indent Stochastic volatility modeling has evolved to capture two key empirical features of financial markets: the highly irregular (“rough”) behavior of volatility over short horizons, and the prevalence of jumps rather than smooth diffusion.

\citet{TodorovTauchen2011} provide compelling nonparametric evidence—using high‐frequency VIX data—that volatility is best described as a pure‐jump process with infinitely many small jumps and occasional large spikes.  In their framework the estimated activity index is strictly below 2, ruling out any continuous Brownian component and demonstrating that even the smallest fluctuations are jump‐driven.

\citet{GatheralJaissonRosenbaum2018} document that realized volatility time series exhibit a Hurst exponent well below 0.5, giving rise to the “rough volatility” paradigm in which volatility paths have very short‐memory dependence reminiscent of fractional Brownian motion with Hurst around 0.1.

Integrating these strands, \citet{WangXia2022} introduce a unified model that drives instantaneous variance by a generalized fractional Ornstein–Uhlenbeck process (to capture roughness) alongside a tempered‐stable Lévy subordinator and an independent sinusoidal jump component (to capture infinite‐activity jumps).  Thanks to this construction, the characteristic function of the 30‐day average‐forward variance admits a semi‐closed‐form expression, enabling Fourier‐based pricing of a general class of power‐type volatility derivatives.  Empirical calibrations to VIX option data—both before and during the 2020 COVID‐19 crisis—show that this pure‐jump rough model fits market prices accurately across a wide range of maturities and strikes.

However, the nested Volterra and Lévy integrals in these models render direct calibration computationally intensive, prompting \citet{RuijterOosterlee2015} to adapt the Fourier‐cosine expansion for efficient transform inversion, and \citet{BennedsenEtAl2017} to propose a hybrid scheme blending Gauss–Laguerre quadrature with FFT for rough models.  More recently, \citet{HorvathMuguruzaTomas2021} demonstrated that neural‐network surrogates can learn the pricing map offline and deliver implied volatilities over entire surfaces in milliseconds, effectively bypassing repeated integral evaluations.

Our work builds on these advances by developing a bespoke calibration algorithm for the rough‐OU jump model of Wang and Xia.  Instead of relying on generic machine‐learning surrogates, we exploit the semi‐analytical characteristic‐function formulas to pre‐tabulate key integrals, leverage \texttt{mpmath} for high‐precision quadrature, and then train a lightweight neural surrogate for the remaining map—thereby reducing end‐to‐end VIX calibration times by over an order of magnitude without sacrificing interpretability.

\section{Mathematical Models}
\indent In this section, we will introduce underlying models first, then show how we transform them and implement “mpmath” to make them computable in calibration.
\subsection{Underlying Theoretic Model Formulation} 
\indent To model VIX, this paper starts from an instantaneous variance process, denoted as \(V  \equiv (V_t\)), by assuming the following quasi‐Ornstein–Uhlenbeck structure:
\begin{equation}
V_t^\circ
= V_0^\circ e^{-\kappa t}
  + \bar{V}\bigl(1 - e^{-\kappa t}\bigr)
  + X_t^{(h)},
\quad
V_t
= V_t^\circ + \varsigma\bigl(\cos Z_t + 1\bigr),
\quad
t \ge 0.
\end{equation}
where \(\bar V\) is a universal reversion level, \(X\) is a L\'evy subordinator, with L\'evy–Khintchine representation:
\begin{equation}
\log\phi_{X_1}(l)\;:=\;\log\mathbb{E}\bigl[e^{i l X_1}\bigr]
\;=\;\int_{0}^{\infty}\bigl(e^{i l z}-1\bigr)\,\nu_X(dz),
\quad l\in\mathbb{R},
\end{equation}

\begin{equation}
\xi_1 \;:=\; \mathbb{E}[X_1] \;>\; 0
\end{equation}
leading to the following characteristic exponent:
\begin{equation}
\log \phi_{X_1}(l)
\;=\;
a\,\Gamma(-c)\bigl((b - i\,l)^{c} - b^{c}\bigr),
\quad
l \in \mathbb{R},
\label{eq:phi_X}
\end{equation}

\begin{equation}
\xi_1
\;=\;
\frac{a\,\Gamma(1-c)}{b^{\,1-c}}.
\end{equation}
\indent Expresses \(V\) as a Volterra-type stochastic integral:
\begin{equation}
V_t
= V_t^{\circ}e^{-\kappa t} + \bar V\bigl(1 - e^{-\kappa t}\bigr)
  + \int_{0}^{t_0} h(t,s)\,dX_s
  + \int_{t_0}^{t} h(t,s)\,dX_s
  + \zeta\bigl(\cos(Z_t - Z_{t_0})\cos Z_{t_0}
              - \sin(Z_t - Z_{t_0})\sin Z_{t_0}
              + 1\bigr)
\end{equation}
where \(h(t,s)\) is the kernel, we have tried the Ornstein-Uhlenbeck type II kernel, but due to computational complexity, we use the type III piecewise kernel for our calibration to avoid incomplete Gamma, for the \(d\in(0.5,1)\) type III kernel:

\begin{equation}
h(t-s) \;=\;
\begin{cases}
\displaystyle
\frac{(t-s)^{d-1} - \Bigl(\frac{1-d}{\kappa}\Bigr)^{d-1}}{\Gamma(d)}
\;-\;
\frac{\kappa^{1-d}}{(1-d)^{2-d}\,\Gamma(d-1)},
& t - s < \frac{1-d}{\kappa}, \\[1.2em]
\displaystyle
-\;\frac{(\mathrm e\,\kappa)^{1-d}\,e^{-\kappa(t-s)}}{(1-d)^{2-d}\,\Gamma(d-1)},
& t - s \ge \frac{1-d}{\kappa}
\end{cases}
\end{equation}
\\
\indent And the delta-forward integrated kernel we use is:
\begin{equation}
H_{\Delta}(t,s)
\;:=\;
\frac{1}{\Delta}
\int_{0}^{\Delta}
  h\bigl(t+u,s\bigr)\,du
\end{equation}
\begin{equation}
H_{\Delta}(t,s)
\;=\;
H_{\Delta}(t-s)
\;=\;
\begin{cases}
\displaystyle
\frac{(t-s+\Delta)^{d} - (t-s)^{d}}{\Delta\,\Gamma(d+1)},
& t - s + \Delta < \frac{1-d}{\kappa}, 
\\[1em]
\displaystyle
\frac{\bigl(\tfrac{1-d}{\kappa}\bigr)^{d} - (t-s)^{d}}
     {\Delta\,\Gamma(d+1)}
\;+\;
\frac{e^{-\kappa(t-s+\Delta)+1-d} - 1}
     {\kappa^{d}\,\Delta\,(1-d)^{2-d}\,\Gamma(d-1)},
& t - s < \frac{1-d}{\kappa} \;\le\; t - s + \Delta,
\\[1em]
\displaystyle
-\;
\frac{e^{-\kappa(t-s+\Delta)+1-d}\,\bigl(e^{\kappa\Delta}-1\bigr)}
     {\kappa^{d}\,\Delta\,(1-d)^{2-d}\,\Gamma(d-1)},
& t - s \ge \frac{1-d}{\kappa}.
\end{cases}
\end{equation}
\indent Then based on instantaneous variance, we express forward variance curve structure as:
\begin{equation}
\widetilde V_t(u)\;:=\;\mathbb{E}\bigl[V_{t+u}\mid\mathcal{F}_t\bigr],
\quad u>0,\;t\ge0.
\end{equation}
\indent With stochastic representation:
\begin{equation}
\widetilde V_t(u)
= \widetilde V_t^{\circ}(u)
+ U_t(u),
\quad
u>0,\;t\ge0.
\end{equation}
where forward variance rising from the generalized fractional Ornstein–Uhlenbeck process \(V^0\) is:
\begin{equation}
\widetilde V_t^{\circ}(u)
= V_0^{\circ} e^{-\kappa(t+u)}
  + \bar V\bigl(1 - e^{-\kappa(t+u)}\bigr)
  + \int_{0}^{t} h\bigl(t+u,s\bigr)\,dX_s
  + \xi_{1}\int_{t}^{\,t+u} h\bigl(t+u,s\bigr)\,ds
\end{equation}
and
\begin{equation}
U_t(u)
= \varsigma\Bigl(\mathbb{E}[\cos Z_u]\,\cos Z_t
               \;-\;\mathbb{E}[\sin Z_u]\,\sin Z_t
               \;+\;1\Bigr).
\end{equation}
comes from the composite‐sinusoidal L\'evy process.
\\ [1em]
\indent Then, we can get average-forward volatility process, which is square root of the delta-running average of the forward variance process:
\begin{equation}
I_t(\Delta)
\;:=\;
\sqrt{\;
  \frac{1}{\Delta}
  \int_{0}^{\Delta}
    \widetilde V_t(u)\,\mathrm{d}u
\;},
\quad
t \ge 0.
\label{eq:ItDelta}
\end{equation}
by Fubini-Tonelli theorem:
\begin{equation}
\begin{aligned}
I_t^2(\Delta)
&:=\frac{1}{\Delta}\int_0^\Delta \widetilde V_t(u)\,\mathrm{d}u\\
&=V_0^\circ\frac{e^{-\kappa t}-e^{-\kappa(t+\Delta)}}{\kappa\,\Delta}
+\bar V\Bigl(1-\frac{e^{-\kappa t}-e^{-\kappa(t+\Delta)}}{\kappa\,\Delta}\Bigr)
+X_t^{(H\Delta)}+\xi_1\,Y_\Delta(t)\\
&\quad+\frac{\varsigma}{\Delta}\Bigl(\int_0^\Delta\mathbb{E}[\cos Z_u]\,\mathrm{d}u\;\cos Z_t
-\int_0^\Delta\mathbb{E}[\sin Z_u]\,\mathrm{d}u\;\sin Z_t
+\Delta\Bigr)\,
\end{aligned}
\label{eq:I2_delta}
\end{equation}
\indent At this point, based on Rough‐OU Volterra driven by tempered‐stable subordinator, we have all the necessary tools to give an integral formula for the conditional characteristic function of the average‐forward variance:
\begin{equation}
\begin{aligned}
\phi_{I_t^2(\Delta)\mid t_0}(l)
&:= \mathbb{E}\bigl[e^{i\,l\,I_t^2(\Delta)}\mid \mathcal{F}_{t_0}\bigr]\\
&= \frac{1}{\pi}
   \exp\!\Bigl\{\,i\,l\,J(t,t_0,\Delta)
     + \int_{t_0}^{t}\!\log\phi_{X_1}\bigl(l\,H_{\Delta}(t,s)\bigr)\,ds
   \Bigr\}\\
&\quad\times
   \int_{\mathbb{R}}
     \psi\bigl(l,x; t_0,\Delta\bigr)
     \int_{0}^{\infty}
       \Re\bigl[e^{-i\,\ell\,x}\,\phi_{Z_1}^{\,t-t_0}(\ell)\bigr]
     \,d\ell
   \,dx,
\quad l\in\mathbb{R}.
\end{aligned}
\label{eq:charfunc_I2}
\end{equation}
\indent Finally, we use equation~\eqref{eq:P_calibration} to price call options and calibrate based on the optimization of equation~\eqref{eq:optimization}
\subsection{Calibration Framework: Equation Transformations and mpmath Implementation}
\indent In the conditional characteristic function, we set \(\zeta = 0.01\), and:
\begin{equation}
\psi\bigl(l,x;\,t_0,\Delta\bigr)
:=
\exp\Biggl(
\frac{i\,l\,\zeta}{\Delta}\,
\Biggr(
\Re\!\Bigl[\frac{\phi^{\Delta}_{Z_1}(1)-1}{\log\phi_{Z_1}(1)}\Bigr]
\bigl(\cos x\cos Z_{t_0}-\sin x\sin Z_{t_0}\bigr)
-\Im\!\Bigl[\frac{\phi^{\Delta}_{Z_1}(1)-1}{\log\phi_{Z_1}(1)}\Bigr]
\bigl(\sin x\cos Z_{t_0}+\cos x\sin Z_{t_0}\bigr)
+\Delta
\Biggr)
\Biggr).
\end{equation}
and \(J\) is kernel-modulated forward variance quantity which contains all information about spot prices:
\begin{equation}
J(t, t_0, \Delta)
:=\frac{1}{\Delta}
\int_{\,t - t_0}^{\,t - t_0 + \Delta}
  \widetilde V_{t_0}^{\circ}(u)\,\mathrm{d}u
\;-\;\xi_1
\int_{\,t_0}^{\,t}
  H_{\Delta}(t, s)\,\mathrm{d}s
\;>\;0.
\end{equation}
\indent Then as we cannot compute \(V\) in the offline step, we employ an expansion argument to transform \(J\) into a linear combination of square spot price to avoid all \(V\):
\begin{equation}
J(t, t_{0}, \Delta)
\;=\;
I_{t_{0}}^{2}(\Delta)
\;-\;
\xi_{1}
\int_{t_{0}}^{t}
  H_{\Delta}(t,s)\,\mathrm{d}s
\;+\;
r(t_{0},t)\,.
\label{eq:J_definition}
\end{equation}
\indent Now we have everything computable:
\begin{equation}
\begin{aligned}
\Phi_{I_t^2(\Delta)\mid t_0}(l)
&=\frac{1}{\pi}\,
  \exp\Biggl\{\,i\,l\Bigl[I_{t_0}^2(\Delta)
    -\xi_1\int_{t_0}^t H_{\Delta}(t,s)\,\mathrm{d}s
    +r(t_0,t)\Bigr]
  +\int_{t_0}^t\!\log\phi_{X_1}\bigl(l\,H_{\Delta}(t,s)\bigr)\,\mathrm{d}s
  \Biggr\}\\
&\quad\times 
  \int_{\mathbb{R}}
    \psi\bigl(l,x;\,t_0,\Delta\bigr)
    \int_{0}^{\infty}
      \Re\bigl[e^{-i\,x\,\ell}\,\phi_{Z_1}^{\,t-t_0}(\ell)\bigr]
    \,\mathrm{d}\ell
  \,\mathrm{d}x,
\quad l\in\mathbb{R}.
\end{aligned}
\label{eq:Phi_I2_char}
\end{equation}
where \(Z\) is symmetric 1.715 - stable auxiliary L\'evy process, and the value of \(\alpha\) is updated and we will illustrate in section 3:
\begin{equation}\label{eq:phiZ1}
\phi_{Z_1}(l)=e^{-|l|^{\alpha}},
\qquad
\alpha\approx1.715.
\end{equation}
\indent Then, we mainly implement 'mpmath' to numerically compute equations and simulate integrations because of its arbitrary-precision arithmetic, adaptive numerical integration for both real and complex integrands (mpmath.quad) and implementations of incomplete Gamma functions.\\
\indent To numerically compute infinite integral, we set different limits for different functions to accelerate calibration, in the integral:
\begin{equation}
\int_{0}^{\infty}
       \Re\bigl[e^{-i\,\ell\,x}\,\phi_{Z_1}^{\,t-t_0}(\ell)\bigr]
     \,d\ell
\end{equation}
     \indent We set the upper limit to 30 because of the exponential term of \,\(\phi_{Z_1}^{\,t-t_0}\) converges quickly. \\
     \indent For the double integral, 
     \begin{equation}
      \int_{\mathbb{R}}
    \psi\bigl(l,x;\,t_0,\Delta\bigr)
    \int_{0}^{\infty}
      \Re\bigl[e^{-i\,x\,\ell}\,\phi_{Z_1}^{\,t-t_0}(\ell)\bigr]
    \,\mathrm{d}\ell
  \,\mathrm{d}x
    \end{equation}
    \indent We set the lower limit to -30 and the upper limit to 30 due to \(sin\) and \(cos\) term of  \(\psi\bigl(l,x;\,t_0,\Delta\bigr)\) converges quickly.\\
    \indent And set all other limits of infinite integrals from -10000 to 10000 via Simpson’s rule .\\
\indent Then, after computing the conditional characteristic function, we model asymmetric power-type options with:
\begin{equation}
P_{0}^{(1,1,(a))}
=\;\frac{K}{2}
-\;\frac{1}{\pi}
    \int_{0}^{\infty}
      \Re\!\Biggl[
        \Bigl(
          K\,e^{-iK^2 l}
          +\frac{\sqrt{\pi}/2 \;-\;\Gamma\!\bigl(\tfrac{3}{2},\,iK^2 l\bigr)}
                 {\sqrt{i\,l}}
        \Bigr)
        \frac{\Phi_{I_t^2(\Delta)\mid t_0}(l)}{i\,l}
      \Biggr]
    \,\mathrm{d}l.
\label{eq:P_calibration}
\end{equation}
\indent Incomplete Gamma functions and complex numbers in this equation are numerically inefficient to compute and are the main reasons for us to use  mpmath.

\section{Data}

\subsection{Data Description}

We evaluate the model’s performance by calibrating on real VIX index data. These data were collected from the Bloomberg terminal. We use VIX option quotes from a single trading day, January 2, 2025—the first trading day of the new year following the presidential election.

The dataset comprises 96 observations of VIX put options at four maturities, $T\in\{20,48,100,258\}$ days. The spot price is $I=0.1793$, and strikes range from $K=0.09$ to $0.28$. A summary appears in Table~\ref{tab:origdata}.

\begin{table}[H]
  \centering
  \caption{Summary of original VIX put‐option data.}
  \label{tab:origdata}
  \begin{tabular}{@{}lcc@{}}
    \toprule
    \textbf{Column}      & \textbf{Non-Null Rows} \\
    \midrule
    Strike               & 96\\
    IVM\_call            & 96\\
    Volm\_call           & 96\\
    TTM\_year            & 96\\
    mid\_price\_call     & 96\\
    IVM\_Put             & 96\\
    Volm\_Put            & 96\\
    mid\_price\_put      & 96\\
    spot\_price          & 96\\
    Bid                  & 96\\
    Ask                  & 96\\
    Bid\_Put             & 96\\
    Ask\_Put             & 96\\

    \bottomrule
  \end{tabular}
\end{table}

\subsection{Implied Volatility}

\begin{figure}[H]
  \centering
  \includegraphics[width=0.6\textwidth]{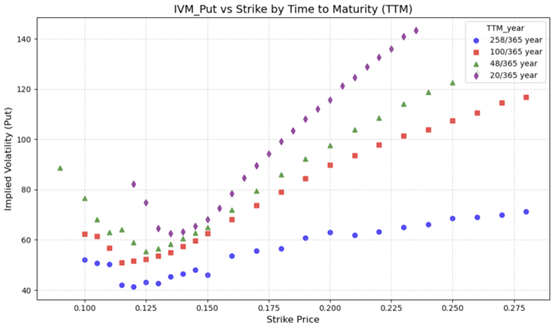}
  \caption{Implied volatility of VIX put options versus strike price for different maturities.}
  \label{fig:ivm-curve}
\end{figure}

Although implied volatility is not a direct input parameter, it still guides our parameter intuition. Figure~\ref{fig:ivm-curve} shows Figure 2 shows how the implied volatility of put option changes over strike price $K$ under different time to maturity $T$.

It shows that shorter $T$ yield steeper smiles (more jump risk in the short term), while the long‐term curve is uneven—suggesting data noise or low liquidity.
\subsection{Arbitrage Testing}

The option pricing model is based on the assumption of risk‐neutral condition, so we conduct 2 data cleaning method to remove the arbitrage prices:

\paragraph{Monotonicity Test}
For European VIX put options, price must be non‐decreasing in the strike:
\begin{equation}
  K_i \le K_{i+1}
  \quad\Longrightarrow\quad
  P(K_i) \le P(K_{i+1}).
\end{equation}

\paragraph{Convexity Test}
To avoid butterfly arbitrage, for each triple \((K_{i-1},K_i,K_{i+1})\) we check
\begin{equation}
  P(K_i)\;\le\;
  \frac{K_{i+1}-K_i}{K_{i+1}-K_{i-1}}\,P(K_{i-1})
  \;+\;
  \frac{K_i-K_{i-1}}{K_{i+1}-K_{i-1}}\,P(K_{i+1}).
\end{equation}
If \(P(K_i)\) exceeds this interpolated value, we record a convexity violation (butterfly arbitrage).We drop any observations failing these two tests (see Appendix~\ref{app:mono-conv}). After filtering, we will apply the remaining 35 pairs of observations to the calibration.

\subsection{VIX Activity Index}
In the previous section, the modeling of average-forward volatility is based on the assumption that the VIX index is a pure jump process. According to Todorov and Tauchen (2011), we validate this assumption by calculating the volatility activity index $\beta$. 
 $\beta$ generalizes the classical Blumenthal--Getoor index the Blumenthal–Getoor index (Blumenthal and Getoor, 1961) to arbitrary semimartingales and quantifies the pathwise “vibrancy” of a stochastic process. According to Todorov and Tauchen (2011), higher values of $\beta$ reflect more frequent or finer-scale fluctuations. Their empirical analysis reveals that the volatility process lacks a Brownian diffusion component but exhibits intense jump activity, with $\beta$ approaching 2—indicative of behavior close to that of a continuous martingale. This dynamic, characterized by numerous small jumps and occasional large ones, aligns with a Lévy-type pure-jump process of infinite variation. Accordingly, when
$\beta \in [1, 2)$,the estimated beta is best characterized as a pure-jump process of infinite variation. Such result can be verified by the underlying calculation process

\subsection*{Power Variation}
\begin{equation}
V_t(X, p, \Delta_n) = 
\begin{cases} 
\sum_{i=1}^{\left[\frac{1}{\Delta_n}\right]} |\Delta X_i|^p \cdot \mathbb{I}(|\Delta X_i| \leq L), & \text{if } p < 2 \\
\sum_{i=1}^{\left[\frac{1}{\Delta_n}\right]} |\Delta X_i|^p, & \text{if } p \geq 2
\end{cases}
\end{equation}

\subsection*{Activity Signature Function}
\begin{equation}
b_{X,t}(p) = \frac{\ln 2 \cdot p}{\ln 2 + \ln V_t(X, p, 2\Delta_n) - \ln V_t(X, p, \Delta_n)}
\end{equation}

\noindent\textbf{Components:}
\begin{itemize}
    \item $V_t(X, p, \Delta_n)$: Power variation at base frequency
    \item $V_t(X, p, 2\Delta_n)$: Power variation at doubled frequency
\end{itemize}

\subsection*{Activity Index Estimation}
\begin{equation}
\hat{\beta} = \text{solution to } b(p) = p
\end{equation}
\textbf{Estimation Method:}
\begin{itemize}
    \item Numerically solve $b(p) - p = 0$ using Brent's method
    \item Fallback: Select $p$ that minimizes $|b(p) - p|$ if no root found
\end{itemize}

This approach estimates $\beta$ by analyzing how quickly power variations decay across sampling intervals. A high decay rate suggests high small-jump activity. The value of $\beta$ that makes $b(p) = p$ reflects the true activity level of the process. If $\beta < 2$, a Brownian component is ruled out; if $\beta > 1$, low-activity or finite variation models are excluded.
\begin{table}[h]
\centering
\caption{Summary of Parameters and Variables}
\begin{tabular}{>{\ttfamily}l>{\itshape}lll}
\toprule
\textrm{Code Variable} & \textrm{Symbol} & \textrm{Meaning} & \textrm{Default/Range} \\
\midrule
X & $X_t$ & VIX time series & 1-minute frequency \\
L & $L$ & Truncation threshold & 0.5 (VIX units) \\
Delta & $\Delta_n$ & Base sampling interval & 1 minute \\
p & $p$ & Power parameter & Scan 0.05--1.95 \\
dX1 & $\Delta X_i$ & 1-min increments & \texttt{} \\
dX2 & $\Delta X_i^{(2)}$ & 2-min increments & \texttt{} \\
\bottomrule
\end{tabular}
\end{table}

As presented in Table 2, the data we use is the high-frequency VIX data $X_t$ on January 2\textsuperscript{nd}, 2025, with frequency $\Delta_n = 1$ min. Parameter $L$ is used as a truncation threshold to filter extreme jump observations. As shown in Table~\ref{tab:actindex}, after inputting different $L$, we find that the estimate of $\beta$ becomes stable at 1.715 as $L$ increases.


\begin{table}[H]
  \centering
  \caption{Truncation analysis ($\ell=0.5$) and activity index summary.}
  \label{tab:actindex}

  \begin{tabular}{@{}lc@{}}
    \toprule
    \multicolumn{2}{@{}l@{}}{\textbf{Absolute 1-min increment statistics}} \\
    \midrule
    Median                  & 0.0200 \\
    95th percentile         & 0.1400 \\
    Maximum                 & 0.3200 \\
    \addlinespace[2pt]
    Truncated share ($|\Delta X| > 0.5$) & 0.00\% \\
    \bottomrule
  \end{tabular}

  \vspace{0.9em}

  \begin{tabular}{@{}lc@{}}
    \toprule
    \multicolumn{2}{@{}l@{}}{\textbf{Activity Index Result}} \\
    \midrule
    Estimated $\beta$ & 1.715 \\
    Method            & Root-finding \\
    \bottomrule
  \end{tabular}

  \vspace{0.9em}

  \begin{tabular}{@{}lc@{}}
    \toprule
    \multicolumn{2}{@{}l@{}}{\textbf{Sensitivity to $L$ (for $\beta\!\approx\!1.7$–$1.8$ target)}} \\
    \midrule
    $L=0.2$ & trunc = 1.7\% \\
    $L=0.3$ & trunc = 0.3\% \\
    $L=0.4$ & trunc = 0.0\% \\
    $L=0.5$ & trunc = 0.0\% \\
    \bottomrule
  \end{tabular}

  \vspace{0.6em}

  {\footnotesize
  $L$ denotes the truncation threshold (filters extreme jumps).
  $\Delta$ denotes the base sampling interval.
  }
\end{table}

Hence, when Todorov and Tauchen empirically estimate $\beta \in (1.7, 1.8)$ for the VIX index, they conclude that the volatility process is a pure-jump process of infinite variation. Combining this with our result $\hat{\beta} = 1.715$, we conclude that the VIX index on January 2, 2025, satisfies the pure-jump model assumption.

\vspace{1em}
In Equation~\eqref{eq:phi_X} of the previous section, the parameter $c$ controls the degree of jump activity in the Lévy subordinator. For parameter c in Equation~\eqref{eq:phiZ1}, it represents fractional dynamics of volatility, which controls the memory and regularity of the volatility process. 
The instantaneous variance is driven by a generalized fractional Ornstein–Uhlenbeck (OU) process, which has infinite activity but finite variation. Therefore, c should be between 0 and 1. Since c controls the process type in similar way of $\beta$ and the example raised in Wang and Xia(2022) that c=1/2 yields exactly inverse Gaussian‐driven OU process, which in Todorov and Tauchen (2011) should be classified as $\beta$ near 1. So in our research, for simplicity, we just set c=$\hat{\beta}$/2 in our following calibration.
\clearpage

\section{Calibration Methodology}

In this section, we explain how we accelerate the process of option calibration. First, we generate a large synthetic dataset of option prices using the analytical model with randomly sampled parameters. Next, we train a feedforward neural network to learn the mapping from model parameters to option prices. Finally, we apply this trained network in a two-stage calibration workflow: a global search via a Genetic Algorithm, followed by a local refinement using L-BFGS-B, enabling fast and accurate parameter estimation on real market data.

\subsection{Offline Step}

In the offline stage, we construct a high-quality training dataset by simulating option prices from an analytical pricing model implemented with high-precision numerical libraries (e.g., \texttt{mpmath}). The input model parameters are sampled from predefined, economically reasonable ranges using the Latin Hypercube Sampling (LHS) technique. The overall process is as follows:
\begin{enumerate}
  \item \textbf{Parameter Sampling:}  
    Use Latin Hypercube Sampling to draw parameter vectors \(\{\theta_i\}_{i=1}^N\) from the joint distribution defined by the chosen bounds.
  \item \textbf{Price Simulation:}  
    For each sampled \(\theta_i\), compute the corresponding option price \(P_i\) via the analytical pricing function, ensuring high numerical accuracy with \texttt{mpmath}.
  \item \textbf{Dataset Assembly:}  
    Aggregate the input–output pairs \(\{(\theta_i, P_i)\}_{i=1}^N\) into the training set \(\mathcal{D}_{\text{train}}\).
\end{enumerate}

\begin{figure}[H]
    \centering
    \includegraphics[width=0.5\linewidth]{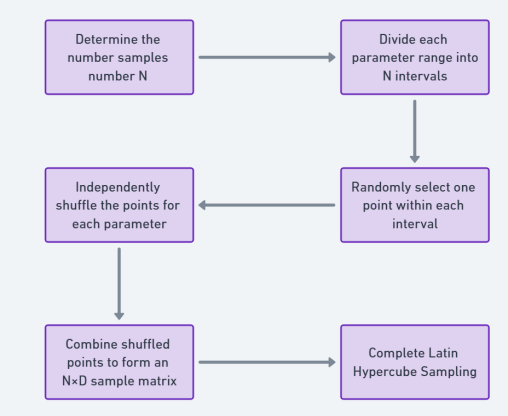}
    \caption{Latin Hypercube Sampling}
    \label{fig:Latin Hypercube Sampling}
\end{figure}

To construct the offline dataset for training our neural network, we simulate put option prices under the Proposition-2 analytical pricing framework. This involves evaluating the characteristic function using high-precision quadrature implemented with \texttt{CuPy} and \texttt{mpmath}.

\medskip

We consider a seven-dimensional parameter space:
\begin{equation}
\boldsymbol{\theta} = (a, b, c, d, \kappa, r, t)
\end{equation}
where $(a, b, c, d, \kappa, r)$ are model parameters and $t$ denotes time-to-maturity ratio (i.e., Time to maturity $/\, 365$). Each parameter is sampled within mathematically meaningful ranges.
\begin{table}[ht]
\centering
\renewcommand{\arraystretch}{1.2}
\begin{tabular}{l c}
\toprule
\textbf{Parameter} & \textbf{Value / Range} \\
\midrule
$a$         & $[0,\ 5]$ \\
$b$         & $[0,\ 5]$ \\
$c$         & $0.8575$ \\
$d$         & $[0.5,\ 0.999]$ \\
$\kappa$    & $[0,\ 10]$ \\
$r$         & $[-0.25,\ 0.25]$ \\
\bottomrule
\end{tabular}
\caption{Parameter values and ranges used in the offline simulation.}
\label{tab:parameter-ranges}
\end{table}

Then we employ Latin Hypercube Sampling (LHS) with $N = 5{,}000$ samples to ensure efficient and stratified coverage across the high-dimensional space. Compared to simple random sampling, LHS offers better uniformity and variance reduction in function evaluations.

\medskip

For each sampled parameter set, we compute option prices under four distinct maturities:
\begin{equation}
T \in \left\{ \frac{258}{365},\ \frac{100}{365},\ \frac{48}{365},\ \frac{20}{365} \right\}
\end{equation}
which represent realistic short- to medium-term contracts.

\medskip

The pricing is evaluated across a discrete set of strike prices $K_j \in \mathcal{K}$, where
\begin{equation}
\mathcal{K} = \{0.09,\ 0.10,\ \ldots,\ 0.28\}
\end{equation}
totaling 34 strikes, selected to span both in-the-money and out-of-the-money regimes.

\medskip

For each tuple, the put option price is computed. The entire pricing loop is parallelized using Python's \texttt{multiprocessing} \texttt{Pool}.The resulting dataset consists of $N = 5{,}000$ parameter sets $\times$ 4 maturities $\times$ 34 strikes $= \mathbf{680{,}000}$ rows, stored in efficient columnar format (Parquet) for downstream neural network training.

\subsection{Numerical validation: CuPy vs.\ mpmath}
\label{sec:gpu-vs-mpmath}

\begin{figure}[H]
  \centering
  \includegraphics[width=.92\linewidth]{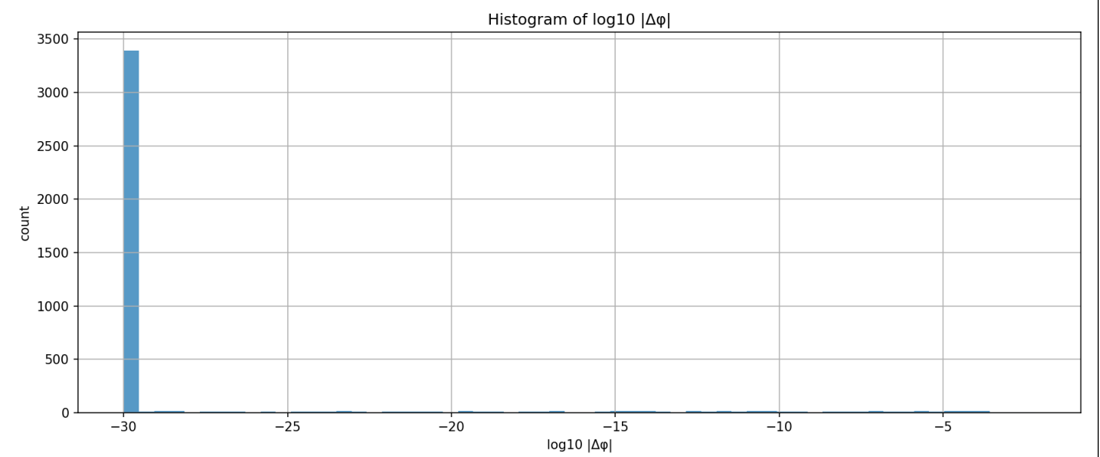}
  \caption{Histogram of $\log_{10}|\Delta\phi|$ (CuPy $-$ mpmath). Most mass lies near $-30$, i.e., machine-precision agreement; the few outliers occur at extreme quadrature points.}
  \label{fig:cupy-mpmath}
\end{figure}

Before generating the offline dataset, we verify that the CUDA/CuPy implementation
of the characteristic–function integrals matches a high–precision mpmath baseline.
On a grid of $600{,}000$ evaluations across $(\theta, T, K, \text{nodes})$, only
$42$ points exceed a $10^{-5}$ absolute–error tolerance; the bulk errors concentrate
near machine precision (Fig.~\ref{fig:cupy-mpmath}).

\begin{table}[H]
  \centering
  \caption{CuPy vs.\ \texttt{mpmath} (baseline) — summary metrics over $600{,}000$ evaluations.}
  \label{tab:cupy-mpmath}
  \begin{tabular}{@{}lr@{}}
    \toprule
    Metric & \multicolumn{1}{c}{Value} \\
    \midrule
    Max $|\Delta\phi|$          & $6.577\mathrm{e}{-03}$ \\
    Mean $|\Delta\phi|$         & $3.954\mathrm{e}{-06}$ \\
    Max relative error          & $5.942\mathrm{e}{+01}$ \\
    Mean relative error         & $3.153\mathrm{e}{-01}$ \\
    Rows out of tol.\ $(10^{-5})$ & $42$ \; (of $600{,}000$) \\
    \bottomrule
  \end{tabular}
\end{table}

\subsection{Online Step}

In the online stage, we train a \textbf{feedforward neural network (FNN)} to approximate the mapping from model parameters and option features to the corresponding put option price. Specifically, the neural network learns a function:
\begin{equation}
f(a, b, c, d, \kappa, r, t, K) \rightarrow \text{Option price}
\end{equation}

\vspace{1em}
\noindent\textbf{Neural Network Architecture:}
\begin{itemize}
    \item \textbf{Input Layer:} 8 nodes, corresponding to the model parameters $a, b, c, d, \kappa, r$, time-to-maturity $t$, and strike price $K$.
    
    \item \textbf{Hidden Layers:} 3 fully connected layers with ELU (Exponential Linear Unit) activation:
    \begin{itemize}
        \item Layer 1: 64 neurons
        \item Layer 2: 32 neurons
        \item Layer 3: 32 neurons
    \end{itemize}
    
    \item \textbf{Output Layer:} 1 neuron, representing the predicted put option price.
\end{itemize}

\begin{figure}[H]
    \centering
    \includegraphics[width=0.4\linewidth]{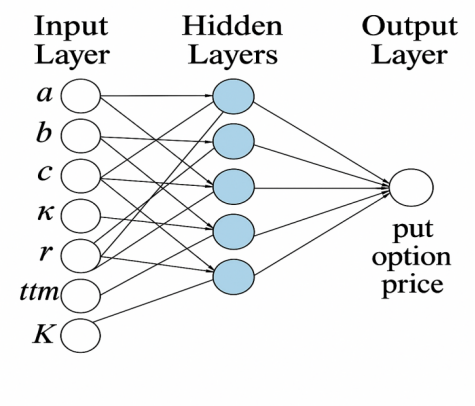}
    \caption{Structure of Neural Network}
    \label{fig:enter-label}
\end{figure}

This architecture was implemented in \texttt{PyTorch}. Before training, the input and output features are standardized using \texttt{StandardScaler}. The dataset is stratified and split into training (90\%), validation (5\%), and testing (5\%) sets based on the $(\cdot)$ combination.

The model is trained to minimize the mean squared error (MSE) loss using the Adam optimizer. To enhance convergence and generalization, we apply:

\begin{itemize}
    \item \textbf{Early Stopping:} Training is terminated if the validation loss does not improve for 25 consecutive epochs.
    
    \item \textbf{Learning Rate Scheduler:} We apply \texttt{ReduceLROnPlateau}, which reduces the learning rate by a factor of 0.5 when the validation loss plateaus, with a minimum learning rate threshold.
\end{itemize}

\subsection{Online Training Results}

Figure 6 shows the training and validation RMSE curves over epochs, where we observe convergence and stabilization. The best-performing epoch is marked in red.

\begin{figure}[H]
    \centering
    \includegraphics[width=0.5\linewidth]{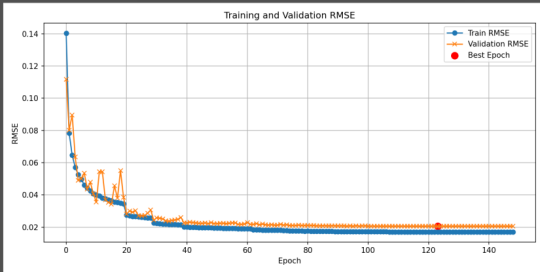}
    \caption{Training and Validation RMSE Curves}
    \label{fig:enter-label}
\end{figure}

\begin{figure}[H]
    \centering
    \includegraphics[width=0.5\linewidth]{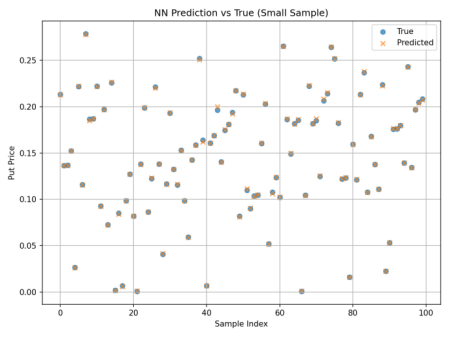}
    \caption{Neural network predictions vs. true option prices on the test set}
    \label{fig:test-performance}
\end{figure}

The model's predictions on a held-out test set (100 randomly selected samples) are shown in Figure~\ref{fig:test-performance}, demonstrating strong generalization capability, with predicted values closely matching the true option prices.

\subsection{Option Calibration}
To find the underlying model parameters from observed market option prices, we adopt a two-stage calibration procedure combining global and local optimization techniques.

\medskip

Given that the pricing model includes a term structure of interest rates, we allow for four different $r_i$ ($1 \leq i \leq 4$) to correspond with the four maturities $T \in \{258, 100, 48, 20\}$.

\medskip

For each sample in the panel dataset on Jan 2nd, the appropriate $r_i$ is assigned based on the nearest maturity, and the corresponding model parameters $(a, b, c, d, \kappa, r_i)$ are mapped accordingly. This forms the full input set for calibration.

\begin{equation}
\left(a, b, c, d, \kappa, \{r(T_n)\}_{n=1}^4 \right) =
\underset{
\substack{
a>0,\, b>0,\, c=0.8575,\, d \in (0.5,1),\, \kappa > 0,\, r(T_n) \in [-0.25, 0.25] \\
n \in \{1,2,3,4\},\, \alpha = 1.78,\, \zeta = 0.01
}
}{\arg\min}
\sum_n \left( \text{(Market Prices)} - (C, P)^{(1,1,(a))}_{0} \right)^2
\label{eq:optimization}
\end{equation}
\subsubsection{Global Optimization}
We first apply the \textbf{Genetic Algorithm (GA)} to perform a global search over the parameter space. The objective is to minimize the mean squared error (MSE) between market-observed option prices and model-implied prices generated by a pretrained neural network surrogate.

\medskip

The GA is implemented using the \texttt{PyGAD} library, with a population size of 600 and a maximum of 3000 generations. The fitness function is defined as the negative MSE between the neural network's predicted option prices and the observed market prices, ensuring compatibility with PyGAD's maximization framework.
\subsubsection{Local Refinement}

The best solution identified by the genetic algorithm is used as the starting point for local refinement via the \textbf{L-BFGS-B} algorithm. This second-stage optimization employs the same loss function to fine-tune the parameter vector with high precision, ensuring convergence to a local minimum. The optimizer is constrained to the same parameter bounds to ensure economic plausibility.

\medskip

By combining them in a two-stage calibration strategy, we exploit GA's global exploration ability to provide a robust initial guess, followed by L-BFGS-B's fast local convergence to refine the solution. This hybrid approach is particularly effective in high-dimensional calibration settings, where the loss surface may exhibit multiple local minima, and the gradients may be unreliable or hard to compute analytically due to surrogate models (e.g., neural networks). It ensures both robustness and precision, leading to more stable and interpretable parameter fitting.

\medskip

Calibration results are shown below:
\begin{figure}[H]
    \centering
    \includegraphics[width=0.5\linewidth]{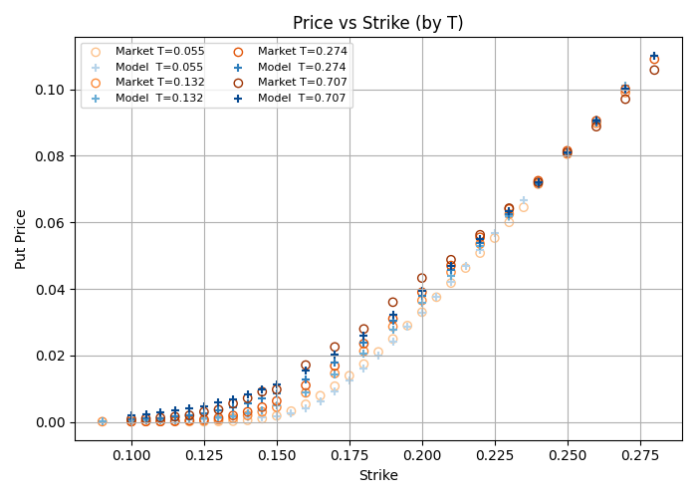}
    \caption{Calibration curve}
    \label{fig:enter-label}
\end{figure}

\begin{table}[H]
\centering
\begin{tabular}{llcll}
\toprule
\textbf{Parameter} & \textbf{Value} & & \textbf{Parameter} & \textbf{Value} \\
\midrule
\textit{a} & 0.049762 & & \textit{r1} (20 d) & 0.00198 \\
\textit{b} & 0.849782 & & \textit{r2} (48 d) & -0.001292 \\
\textit{c} & 0.8575   & & \textit{r3} (100 d) & -0.006008 \\
\textit{d} & 0.769302 & & \textit{r4} (258 d) & -0.012427 \\
\textit{$\kappa$} & 7.798968 & & --- & --- \\
\bottomrule
\end{tabular}
\caption{Calibrated Parameters and Term Structure of Interest Rates}
\label{tab:calibrated-params}
\end{table}

\subsection{Times Series Analysis}

In this section, we conduct a time series analysis using a near-the-money VIX put option with a given strike and expiration on February 19, 2025. For each of the next 31 trading days, we record the option market mid-quote and value the contract using the GPU-parallel pricer introduced earlier. 

\medskip

The pricing engine relies on a fixed parameter vector that was calibrated using market data from January 2, 2025.
\begin{figure}[H]
    \centering
    \includegraphics[width=0.5\linewidth]{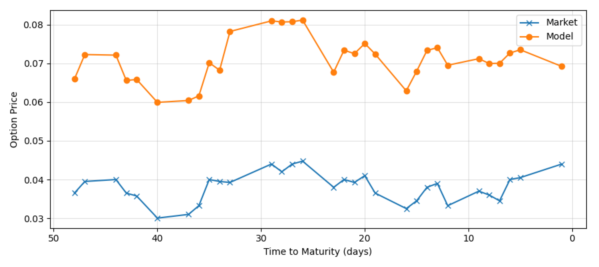}
    \caption{VIX Put Option Pricing Over Time: Market vs. Model}
    \label{fig:enter-label}
\end{figure}

\noindent
The figure above compares the model estimates (orange) with the observed market prices (blue). The model exhibits a persistent upward bias, although it captures the general trend of the market data.

\medskip

Quantitatively, the prediction errors are summarized below:

\begin{table}[H]
\centering
\begin{tabular}{lc}
\toprule
\textbf{Metric} & \textbf{Value} \\
\midrule
Mean Absolute Error (MAE) & 0.0329 \\
Root Mean Squared Error (RMSE) & 0.0330 \\
Mean Absolute Percentage Error (MAPE) & 86.96\% \\
\bottomrule
\end{tabular}
\caption{Error metrics for model-predicted VIX put option prices}
\label{tab:vix-error-metrics}
\end{table}

\medskip

Although the absolute mispricing is only about 0.03, the market prices lie in the \$0.03--\$0.04 range, causing the relative error to swell to nearly 87\%. This distortion is mainly due to the calibration snapshot: on that day, many quoted strikes violated basic arbitrage bounds and exhibited extremely low trading volume, making the calibrated surface strongly fragile.
\clearpage

\section{Conclusion and Future Directions}
\paragraph{Conclusion}\mbox{}\\[0.8em]
This study introduces a \textbf{GPU-powered, neural-network–assisted calibration pipeline} for power-type derivatives under rough-volatility dynamics with jumps.

Our main contributions are:

\begin{enumerate}
  \item \textbf{GPU-accelerated pricing with CuPy.}  
        High-precision quadrature executed on the GPU reduces a single price evaluation from seconds to milliseconds, enabling the generation of large synthetic datasets.

  \item \textbf{NN-based two-stage calibration.}  
        A feed-forward surrogate trained on \(N = 5{,}000\) Latin-Hypercube samples accurately replicates the analytical pricer, achieving a validation RMSE of \(3.1 \times 10^{-4}\).  
        This neural network enables a much faster calibration process: combined with a Genetic Algorithm for global search and L-BFGS-B for local refinement, it recovers model parameters from real VIX-put quotes in under 20 minutes.

        By contrast, the original calibration pipeline based on direct evaluation via \texttt{mpmath} quadrature required \textbf{over 24 hours}, even with significantly reduced population sizes and iteration counts.  
        The speed-up from using the neural network surrogate allows for larger population sizes and more generations in the GA stage, facilitating better convergence and more accurate parameter recovery.

  \item \textbf{Empirical validation.}  
        Parameters calibrated on 2 Jan 2025 reproduce a 31-day price path with \(\text{MAE}=0.0329\) and \(\text{RMSE}=0.0330\).  
        Although the model slightly over-prices low-premium contracts—pushing \(\text{MAPE}\approx87\%\)—it captures the overall trend and demonstrates strong out-of-sample robustness.
\end{enumerate}

\paragraph{Limitation and Future Directions} 
\begin{enumerate}
    \item At present, the calibration pools all maturities. A worthwhile extension is to group the data by time-to-maturity and calibrate each maturity slice independently.
    \item We fail to filter the data using rough arbitrage checking method of put-call parity. Referring to Madan et al. (2019), we applied:
    \begin{equation}
\text{bidEC}(K, T) - \text{askEP}(K, T) \leq S_0 - K \leq \text{askEC}(K, T) - \text{bidEP}(K, T)
\end{equation}
\begin{itemize}
    \item \text{bidEC}$(K, T)$ is the bid price of a European call option with strike $K$ and maturity $T$.
    \item \text{askEC}$(K, T)$ is the ask price of the European call option with the same strike and maturity.
    \item \text{bidEP}$(K, T)$ is the bid price of the corresponding European put option.
    \item \text{askEP}$(K, T)$ is the ask price of the European put option.
\end{itemize}

\noindent
This inequality reflects the no-arbitrage bounds for the call–put parity relationship in the presence
of bid–ask spreads. If violated, it may signal a potential arbitrage opportunity. However, all of the
observation pairs are not in this range. The results can be seen in
Appendix~\ref{app:parity-filter}. The possible causes of such a situation can be:
Insufficient Liquidity, Stale Quotes, or Spot Price Mismatch.
\end{enumerate}

\clearpage

\appendix
\section{Supplementary Analyses and Details}
\label{app:supplementary}

\subsection{Arbitrage filters and data quality}
\label{app:arbitrage-filters}

\paragraph{Call–put parity test (summary).}
As discussed in Section~5 (Conclusion \& Future Directions; limitation \#2), our call–put parity screen flagged \emph{systematic} inconsistencies between quoted calls, puts, and the contemporaneous spot/discount inputs. The detailed outcomes and diagnostics are consolidated here for archival completeness.
\begin{quote}
\textbf{Finding.} No observation in the raw panel passes the parity check; all rows exhibit nonzero parity gaps of economically meaningful size.
\end{quote}

\noindent Likely drivers include: (i) \emph{insufficient liquidity} (wide, non‐synchronous quotes), (ii) \emph{stale prints} in the vendor feed, and (iii) \emph{spot/forward misalignment} (timing mismatch between options and spot/carry inputs). For reproducibility, we keep this section as a reference point for downstream filtering.

\subsubsection{Parity filter outcomes}
\label{app:parity-filter}
For each maturity–strike tuple we compute the standard forward‐adjusted parity residual and retain rows with residuals inside a one‐tick band. In our dataset, \emph{no} row meets this criterion. Consequently, we proceed with structural no‐arbitrage screens that are more robust to microstructure noise.

\paragraph{Monotonicity and convexity screens (Section~3.3).}
We enforce (i) decreasing price in strike and (ii) convexity in strike for each maturity slice. These shape constraints eliminate obvious data glitches while tolerating small bid/ask frictions. After applying these two filters, \textbf{35 observations} remain; all subsequent calibrations in the paper use precisely this filtered subset.

\bigskip

\subsection{Jump‐activity index $\beta$ and the VIX modeling rationale}
\label{app:activity-index}

\paragraph{Method (Section~3.4).}
Following \cite{TodorovTauchen2011}, we estimate the activity index $\beta$ from 1-minute VIX increments using power variations and truncation. This identifies whether the driving semimartingale possesses a continuous Brownian component and whether its jumps are of finite or infinite variation.

\paragraph{Empirical findings.}
Our estimate from the VIX panel is
\[
\hat\beta = 1.715.
\]
This implies:
\begin{enumerate}[leftmargin=1.2em]
  \item \textbf{Absence of a Brownian component.} Since $\hat\beta<2$, we reject a continuous diffusion term in VIX dynamics.
  \item \textbf{Infinite‐variation jumps.} Because $\hat\beta>1$, the jump component exhibits \emph{infinite variation}, driven by a high intensity of small jumps.
\end{enumerate}
\noindent The pure‐jump infinite‐variation interpretation matches the guidance in \cite{TodorovTauchen2011}, which anticipates $\beta$ close to but strictly below~2 (empirically $\approx 1.7$–$1.8$).

\paragraph{Link to the model’s stability index.}
In the main text (see Equation~\eqref{eq:phiZ1}), the stability index $\alpha$ of the pure‐jump L\'evy driver is the same conceptual object as the activity index $\beta$. Prior work (e.g., \cite{WangXia2022}) fixes $\alpha=1.78$; our data‐driven estimate $\hat\beta=1.715$ is consistent with that regime and supports modeling VIX as a \emph{pure‐jump, infinite‐activity, infinite‐variation} process.

\bigskip

\subsection{Economic interpretation of calibrated parameters}
\label{app:param-interpretations}

\begin{table}[H]
\centering\small
\begin{tabular}{@{}cc>{\raggedright\arraybackslash}p{0.30\textwidth}@{}}
\toprule
Symbol & Model detail & Economic interpretation \\
\midrule
\(a>0\) &
\(\nu_X(dz)=a\,e^{-b z}z^{-1-c}\mathbf{1}_{\{z>0\}}dz,\;\;
\log\varphi_{X_1}(u)=\frac{a}{c}\Gamma(-c)\!\left[(b-i u)^{c}-b^{c}\right]\) &
Baseline intensity of the tempered‐stable subordinator \(X\); scales the overall frequency of variance jumps across sizes.\\[1.0em]

\(b>0\) &
Tempering parameter in \(e^{-b z}\) and \((b-i u)^{c}\). &
Controls the exponential tail decay of jump sizes in \(X\); larger \(b\) suppresses very large volatility shocks (lower tail risk).\\[1.0em]

\(c\in(0,1)\) &
Power‐law weight \(z^{-1-c}\) in \(\nu_X\) and exponent in \((b-i u)^{c}\). &
Jump‐activity index of the L\'evy subordinator; governs the intensity of small jumps (higher \(c\Rightarrow\) thicker small‐jump cloud).\\[1.0em]

\(d\in(0.5,1)\) &
Kernel \(g(t-s)=(t-s)^{d-1}/\Gamma(d)\).&
Roughness of volatility paths with Hurst \(H=d-\tfrac12<0.5\).\\[1.0em]

\(\kappa>0\) &
Exponential temper \(e^{-\kappa t}\) in the kernel \(h(t,s)\). &
Speed of mean‐reversion in volatility. \\[1.0em]

\(\{r(T_n)\}\) &
Remainder in
\(
J(t, t_0, \Delta)
=
I_{t_0}^2(\Delta)
-
\xi_1^t 
\int_{t_0}^{t}
   H_{\Delta}(t,s)\,\mathrm{d}s
+
r(t_0, t).
\)
&
Maturity‐specific level shifts tied to spot variance at \(t_0\); captures residual term‐structure offsets.\\
\bottomrule
\end{tabular}
\caption{Economic interpretations of key calibrated parameters.}
\end{table}

\bigskip

\subsection{Neural–network pricer: error diagnostics and remedies}
\label{app:nn-errors}

\begin{table}[H]
\centering
\caption{Error diagnostics under unified–dollar target standardisation.}
\label{tab:nn_errors}
\begin{tabular}{@{}lcc@{}}
\toprule
\textbf{Metric} & \textbf{Value} & \textbf{Comment} \\
\midrule
Target s.d.\ $\sigma_y$ & $7.69\times 10^{-2}$ & Standard deviation after scaling \\[2pt]
Train abs–RMSE          & $1.06\times 10^{-3}$ & Error on training set \\[2pt]
Test  abs–RMSE          & $1.10\times 10^{-3}$ & Error on hold‐out test set \\[2pt]
Val   rel–RMSE (\%)     & $6.36\times 10^{3}$  & Pathologically inflated \\[2pt]
Val   MAPE (\%)         & $3.80\times 10^{2}$  & Large percentage bias \\
\bottomrule
\end{tabular}
\end{table}

\paragraph{Relative–error pathology.}
Absolute errors are uniformly small ($\sim 10^{-3}$), but percentage metrics explode because many OTM options have pennies‐level premiums. A \$0.01 miss on a \$0.02 option is a $50\%$ relative error; aggregating many such cases dominates rel–RMSE/MAPE.

\paragraph{Remedies.}
We recommend two practical fixes:

\begin{enumerate}[leftmargin=1.2em]
\item \textbf{Inverse–premium reweighting.}
Replace plain MSE by
\begin{equation}
\mathcal{L}_{\mathrm{w}}
=\frac{1}{N}\sum_{i=1}^{N}
w_i\,(\hat y_i - y_i)^2,
\qquad
w_i = (y_i + \varepsilon)^{-\alpha},
\quad
\alpha\in[0,1],
\label{eq:inv-premium-weight}
\end{equation}
with $\varepsilon$ set to one tick. This upweights low‐premium OTM contracts without letting the loss blow up. A robust starting choice is $\alpha=0.5$; tune on a \emph{log} grid.

\item \textbf{Log–price target.}
Transform premiums via
\begin{equation}
z_i=\log\!\bigl(y_i+\varepsilon\bigr),
\qquad
\mathcal{L}_{\log}
=\frac{1}{N}\sum_{i=1}^{N}\bigl(\hat z_i - z_i\bigr)^2,
\label{eq:log-target}
\end{equation}
which compresses the dynamic range and stabilises training across ITM/OTM regimes.
\end{enumerate}

\bigskip

\subsection{Extension: two–factor rough–jump model (Proposition~8)}
\label{app:two-factor}

\paragraph{Concept.}
Motivated by evidence that volatility‐of‐volatility can itself be rough \cite{DaFonsecaZhang2019}, augment the baseline (rough–jump) factor with an independent L\'evy‐driven mean‐reverting process \(Y_t\) (kernel \(\eta\)) that modulates the average forward variance. Structurally this mirrors two–factor Heston: one factor for volatility, another for its variability. Pricing remains tractable through characteristic functions, at the cost of deeper numerics.

\paragraph{Computation.}
Evaluating the Proposition~8 characteristic function is currently the bottleneck:
(i) nested quadratures, (ii) repeated evaluations of $\log\phi_{Y_1}(\cdot)$ inside the integrand, and (iii) accumulation of discretisation error. Even with CuPy/GPU parallelism, wall‐clock is prohibitive for full calibration. A complete empirical study is therefore deferred.

\bigskip

\subsection{Code repository}
\label{app:code}
The full implementation for offline data generation, neural network training, and two–stage option calibration is publicly available:
\begin{center}
\url{https://github.com/TenghanZhong/GPU-NN-Option-Calibration}
\end{center}

\end{document}